\RequirePackage[utf8]{inputenc}
\pdfoutput=1
\documentclass[aps,pra,10pt,superscriptaddress,noeprint,twocolumn]{revtex4-2}

\usepackage[T1]{fontenc}
\usepackage{microtype}
\usepackage{printlen}
\usepackage{amsmath,amssymb,amsfonts,amsthm,bbm}
\usepackage{bbold}
\usepackage{braket}
\usepackage{mathtools}

\usepackage{bm}
\usepackage{graphicx}
\usepackage[dvipsnames]{xcolor}
\usepackage{float}
\usepackage{tikz}
\usepackage[caption=false]{subfig}

\colorlet{mylinkcolor}{teal}
\colorlet{mycitecolor}{teal}
\colorlet{myurlcolor}{teal}

\usepackage{hyperref}
\hypersetup{
  linkcolor  = mylinkcolor,
  citecolor  = mycitecolor,
  urlcolor   = myurlcolor,
  colorlinks = true,
  breaklinks = true
}

\newcommand{\I}{\mathrm{i}}

\DeclarePairedDelimiterX{\norm}[1]{\lVert}{\rVert}{#1}
\DeclarePairedDelimiterX{\abs}[1]{\lvert}{\rvert}{#1}

\makeatletter
\let\oldabs\abs
\def\abs{\@ifstar{\oldabs}{\oldabs*}}
\let\oldnorm\norm
\def\norm{\@ifstar{\oldnorm}{\oldnorm*}}
\makeatother

\DeclareMathOperator{\Tr}{Tr}

\let\originalleft\left
\let\originalright\right
\renewcommand{\left}{\mathopen{}\mathclose\bgroup\originalleft}
\renewcommand{\right}{\aftergroup\egroup\originalright}

\begin{document}

\title{Quantum asymmetry and noisy multi-mode interferometry}
\author{Francesco Albarelli}
\affiliation{Faculty of Physics, University of Warsaw, 02-093 Warsaw, Poland}
\affiliation{Dipartimento di Fisica ``Aldo Pontremoli'', Università degli Studi di Milano, via Celoria 16, 20133 Milan, Italy}

\author{Mateusz Mazelanik}
\affiliation{Centre for Quantum Optical Technologies, Centre of New Technologies, University of Warsaw, 02-097 Warsaw, Poland}
\affiliation{Faculty of Physics, University of Warsaw, 02-093 Warsaw, Poland}

\author{Michał Lipka}
\affiliation{Centre for Quantum Optical Technologies, Centre of New Technologies, University of Warsaw, 02-097 Warsaw, Poland}

\author{Alexander Streltsov}
\affiliation{Centre for Quantum Optical Technologies, Centre of New Technologies, University of Warsaw, 02-097 Warsaw, Poland}

\author{Michał Parniak}
\affiliation{Centre for Quantum Optical Technologies, Centre of New Technologies, University of Warsaw, 02-097 Warsaw, Poland}
\affiliation{Niels Bohr Institute, University of Copenhagen, 2100 Copenhagen, Denmark}

\author{Rafał Demkowicz-Dobrzański}
\affiliation{Faculty of Physics, University of Warsaw, 02-093 Warsaw, Poland}

\begin{abstract}
Quantum asymmetry is a physical resource which coincides with the amount of coherence between the eigenspaces of a generator responsible for phase 
encoding in interferometric experiments.
We highlight an apparently counter-intuitive behavior that the asymmetry may \emph{increase} as a result of a \emph{decrease} of coherence inside a degenerate subspace.
We intuitively explain and illustrate the phenomena by performing a three-mode single-photon interferometric experiment, where one arm carries the signal and two noisy reference arms have fluctuating phases.
We show that the source of the observed sensitivity improvement is the reduction of correlations between these fluctuations and comment on the impact of the effect when moving from the single-photon quantum level to the classical regime. Finally, we also establish the analogy of the effect in the case of entanglement resource theory.
\end{abstract}

\maketitle

Superposition of quantum states is a fundamental principle of quantum mechanics and the ability to create and preserve coherent superpositions is the essential prerequisite for the realization of all kinds of quantum technologies.
In recent years, the basic intuition that quantum superposition is a valuable and fragile resource has been formalized within the mathematical framework of quantum resource theories~\cite{Streltsov2016a,Chitambar2018,Wu2021}.
The mathematical notion of coherence relies on the decomposition of the Hilbert space of the system
into subspaces; usually the orthogonal eigenspaces of an observable.
This is actually a subcase of the more general resource theory of quantum asymmetry~\cite{Gour_2008,Marvian2016}, where the resource is the degree to which a quantum state breaks a certain symmetry, defined in terms of a Lie group.
Quantum asymmetry has been recognized as the relevant physical resource in a variety of operational settings: reference frame alignment~\cite{Bartlett2007,Vaccaro2008,Gour2009}, quantum thermodynamic tasks~\cite{Lostaglio2015,Lostaglio2015b,Korzekwa}, quantum speed limits~\cite{Marvian2016b,Mondal2016a}, assessing macroscopic quantumness~\cite{Yadin2016,Kwon2017b,Frowis2018}, and, most importantly for this work, quantum metrology~\cite{Marvian2014a,Marvian2014,Marvian2016,Zhang2017i}.
In this framework, the coherence of a quantum state with respect to the eigenspaces of an observable $G$ corresponds to the asymmetry with respect to the one-parameter group of translations $e^{\I \theta G}$; in the following, the term quantum asymmetry will be used to refer to this specific notion.

In this work, we focus on a qutrit example, which is the simplest case where a non-trivial generator $G$ with a degenerate eigenvalue exists.
Our main result is to show that the sensitivity to the phase $\theta$, which also quantifies the $G$-asymmetry, is increased for probe states that are more dephased in the degenerate subspace, thus effectively ``noisier'', in a sense that will be made more precise.
Furthermore, we also show that a similar, yet less pronounced, behavior appears when the dephasing channel $\mathcal{E}$ is used for entanglement distribution.

We study the phenomena in terms of a physical realization of the system as a three-arm interferometer and we complement our theoretical analysis with a single-photon optical experiment that confirms our predictions.

While quantum coherence theory is intimately related with  multiple phase interferometry~\cite{Durr2001,ENGLERT2011,Humphreys2013, Bera2015,Bagan2016,Biswas2017,Masini2021}, in this work we focus on the estimation of a single phase $\theta$, imprinted onto a signal mode, while the phases of the other reference modes fluctuate, generalizing the paradigmatic phase diffusion model~\cite{Genoni2011a,Genoni2012,Escher2012,Vidrighin2014,Macieszczak2014,Demkowicz-Dobrzanski2015a,Scott2021a}.
Interestingly, this interferometric point of view will shed light on the apparently unintuitive effect observed in the abstract description: we will show that an increased dephasing in the degenerate subspace is caused by less correlated fluctuations of the reference phases, which allows to decrease the measurement noise without altering the signal.

\emph{Asymmetry and the Quantum Fisher Information.}
Quantum resource theories, with the entanglement theory as the most famous example~\cite{HorodeckiRevModPhys.81.865}, arise naturally when a set of quantum states can be regarded as free and any state outside of it as a resource~\cite{Chitambar2018}.
This immediately defines also free operations, i.e. quantum channels that cannot transform free states into resources.
In the resource theory of asymmetry the starting point is the representation of a  symmetry group.
In particular, we consider a finite-dimensional representation of $U(1)$: $U_{\theta}=e^{-\I \theta G}$, where $G=\sum_{k} E_k | k \rangle \langle k |$ is the Hermitian generator, $\ket{k}$ are its eigenstates and $E_k$ its eigenvalues, which can be degenerate.
Resource states $\mathcal{G}$ are those \emph{not} invariant under the action of $U_{\theta}$ and free operations those commuting with $U_{\theta}$.
If $G$ is nondegenerate, $\mathcal{G}$ is the set of incoherent states with no off-diagonal elements in the eigenbasis; otherwise, all superpositions of eigenstates of the same degenerate eigenvalue are also free.
 
There are inequivalent ways to quantify the asymmetry of a quantum state, called asymmetry monotones, which must not increase under free operations~\footnote{More precise definitions can be found elsewhere~\cite{Streltsov2016a,Chitambar2018} and are not needed for our discussion.}.
When $U_{\theta}$ is regarded as a $\theta$-parameter encoding operation on a quantum state, the quantum Fisher information (QFI) of the state is an asymmetry monotone ~\cite{Zhang2017i}.
It is defined as $\mathcal{F}_{\theta}(\rho)=\Tr [ \rho_\theta L_{\theta}^2 ]$ where $\rho_\theta = U_{\theta} \rho U_{\theta}^\dag$ and the symmetric logarithmic derivative (SLD) operator $L_{\theta}$ is defined as $2 \frac{d \rho_\theta}{d \theta} = L_{\theta} \rho_{\theta} + \rho_{\theta} L_{\theta}$.
The QFI quantifies the sensitivity of $\rho$ to the imprinted phase $\theta$, since the quantum Cramér-Rao bound (CRB) states that the variance of any unbiased estimator $\tilde{\theta}$ is lower bounded as~\cite{helstrom1976quantum,Holevo2011b,Paris2009}
$\Delta^2 \tilde{\theta} \geq \frac{1}{\nu \mathcal{F}_{\theta}(\rho)}$,
where $\nu$ is the number of repetitions and the bound is saturable for large $\nu$.

\emph{Qutrit model.}
Consider a three-level system (qutrit) and the generator $G= E_1 |1\rangle\langle1| + E_2 \left( |2\rangle\langle 2| + |3 \rangle \langle 3| \right) $, where we have fixed an orthonormal basis $\{\ket{2},\ket{3}\}$ of the degenerate eigenspace, singled out by the dephasing channel $\mathcal{E}$, which describes a decrease of coherence between the two eigenspaces by a factor $\eta$ 
and between $\{\ket{2},\ket{3}\}$ by a factor $\kappa$:
\begin{equation}
  \label{eq:quditDeph12}
  \mathcal{E} \left( \rho \right)=
  \begin{bmatrix}
  \rho_{11} & \eta \rho_{12} & \eta \rho_{13} \\
  \eta \rho_{21} & \rho_{22} & \kappa \rho_{23} \\
  \eta \rho_{31} & \kappa \rho_{32} & \rho_{33}
  \end{bmatrix}.
\end{equation}
The channel is physical (a completely positive map) provided
\begin{equation}
\label{eq:CPconds}
0 \leq \eta \leq 1 \quad \land \quad  2 \eta^2 - 1 \leq \kappa \leq 1,
\end{equation}
we restrict to real-valued $\eta$ and $\kappa$ without loss of generality.



We consider the phase parameter $\theta$ unitarily encoded by $G$, i.e. $\mathcal{E}_{\theta}(\rho) = e^{-\I \theta G} \mathcal{E}(\rho) e^{\I \theta G}$
and to study the effect of dephasing we focus on a pure probe state $\rho_0=|\psi_0\rangle\langle\psi_0|$ that is a superpostion of states from different eigenspaces: $\ket{\psi_0} = \sqrt{q} \ket{1} + \sqrt{\frac{1-q}{2}} ( \ket{2} + \ket{3})$, with $0\leq q \leq 1$.
The corresponding QFI is 
\begin{equation}
\label{eq:qutritQFIetakappa}
\mathcal{F} \left( \mathcal{E}_{\theta}(\rho_0) \right) = \frac{8 (E_1 - E_2)^2 \eta^2 (1-q) q}{\kappa + 1 + q(1-\kappa)}
\end{equation}
and quantifies both the asymmetry of the state $\mathcal{E}(\rho)$ and its phase sensitivity.
Details on the calculation are in~\cite{sm}.

As expected, the QFI is a monotonically increasing function of $\eta$. 
However, for any value of $\eta$ and $q$ it is a \emph{decreasing} function of $\kappa$.
This means that more dephasing in the degenerate subspace gives higher sensitivity to the imprinted phase $\theta$, despite the state becoming in some sense noisier---decreasing $\kappa$ from $1$ (no dephasing) to a value $\kappa \geq 0$ always decreases its purity $\Tr \rho^2$.
While this result appears counter-intuitive, we stress that it does not violate the monotonicity property of the QFI.
According to~\eqref{eq:CPconds}, a channel 
with $\eta=1$ and $\kappa<1$ is not physical.
Thus, it is not possible to see the two decoherence effects parametrized by $\eta$ and $\kappa$ as the action of separate channels.

Since $\mathcal{E}$ and $e^{-\I \theta G}$ commute, the overall channel $\mathcal{E}_{\theta}$ also describes a physical process in which noise and parameter encoding happen simultaneously.
While the QFI quantifies the asymmetry of the noisy state $\mathcal{E}(\rho)$, from a quantum metrology point of view one assumes that the channel $\mathcal{E}_{\theta}$ is given and optimizes the probe state.
The so-called channel QFI $\max_{ \rho }\mathcal{F}\left(\mathcal{E}_\theta( \rho ) \right)=8 \eta^2 \left(\kappa-2 \sqrt{2} \sqrt{\kappa+1}+3\right)/(\kappa-1)^2$
is obtained for $q=q_{\text{opt}}=1/\left( \sqrt{2/(\kappa+1)} + 1 \right)$;
apart from the $\kappa=1$ case, the optimal state is not a balanced superposition.

Our results are not due to a peculiar behaviour of the QFI; we have also evaluated the relative entropy of asymmetry~\cite{Aberg2006,Vaccaro2008,Gour2009,Hall2012} and observed a qualitatively analogous behaviour, with the maximal asymmetry achieved for a slightly different value of $q$.
However, the simplest and in some sense the most intuitive measure, 
 i.e, the sum of the trace norm of the modes of asymmetry~\cite{Marvian2014a}, 
 does not depend on $\kappa$ and hence would not reveal the effect.

\begin{figure}[t]
\includegraphics[width=1\columnwidth]{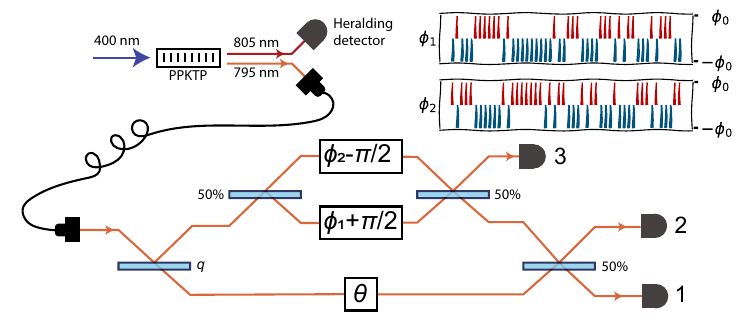}
\caption{Schematics of the experimental demonstration of phase estimation in the three-arm interferometer with correlated noise. Heralded single photons from spontaneous parametric downconversion are used to probe the system with noisy phases $\phi_1$ and $\phi_2$ to estimate the phase $\theta$.
Inset shows example realization of phase noise with high anti-correlation.}
\label{fig:interf_scheme_sing_phot}
\end{figure}

\emph{Interferometric realization.}
We now show a simple physical explanation of the apparently counterintuitive result. Consider a  three-mode interferometer, as depicted in Fig.~\ref{fig:interf_scheme_sing_phot}. 
A single photon is prepared in a three mode superposition by the action of two beam splitters, creating the initial state $\ket{\psi_0}$.
Then the phase $\theta$ is imprinted on the first signal mode, while the second and third reference modes do not have stable phases but are subject to random phase fluctuations $\phi_1$ and $\phi_2$, with zero mean and a joint probability distribution $P(\phi_1,\phi_2)$.
The simplest choice is given by correlated ``phase kicks'', i.e. $\phi_1$ and $\phi_2$ fluctuate to two equal and opposite values $\pm \phi_0 \in [-\frac{\pi}{2},\frac{\pi}{2}]$ with probabilities $P(\phi_0,\phi_0)=P(-\phi_0,-\phi_0)=\frac{1}{4}(1+c)$ and $P(\phi_0,-\phi_0)=P(-\phi_0,\phi_0)=\frac{1}{4}(1-c)$, where $c=\mathbb{E}[\phi_1 \phi_2]/\sqrt{\mathbb{E}[\phi_1^2]\mathbb{E}[\phi_2^2]}$ is the correlation coefficient, ranging from perfectly correlated $c=1$ to perfectly anticorrelated $c=-1$ and $\mathbb{E}$ denotes the expectation with respect to $P(\phi_1,\phi_2)$.
These parameters are related to the previous ones as
\begin{align}
\label{eq:etakappa}
\eta &= \mathbb{E}[ e^{i \phi_1} ] = \mathbb{E}[ e^{i \phi_2} ] = \cos \phi_0  \\
\kappa &= \mathbb{E}[ e^{i (\phi_1-\phi_2) } ] = \cos^2 \phi_0 (1 - c ) + c = \eta^2 + c (1 - \eta^2 ),\nonumber
\end{align}
as the action of the dephasing channel corresponds to  $\mathcal{E} ( \star ) = \mathbb{E}\left[ e^{-\I (\phi_2 |2\rangle \langle 2| + \phi_3 |3\rangle\langle 3|) } \star e^{ \I (\phi_2 |2\rangle \langle 2| + \phi_3 |3\rangle\langle 3| ) } \right]$.
The parameter $\phi_0$ represents the magnitude of the phase kicks and is directly linked to the parameter $\eta$, which we will keep using for convenience.
The parameter $\kappa$ on the other hand, depends both on $\phi_0$ and on the the correlation coefficient $c$, and the whole range of physical values of $\kappa$ in~\eqref{eq:CPconds} can be obtained by varying $c$.
It is also possible to obtain the same qutrit channel $\mathcal{E}$ if $P(\phi_1,\phi_2)$ is a bivariate Gaussian distribution, similarly to the standard phase diffusion model in a Mach-Zehnder (MZ) interferometer~\cite{Genoni2011a,Escher2012,Szczykulska2017}.
The qualitative picture remains the same: the less correlated (i.e. more anti-correlated) $\phi_1$ and $\phi_2$ are, the higher the phase sensitivity is.
However for $c<0$ the effect is less pronounced, since this noise model does not reproduce the full range of physical $\kappa$ in Eq.~\eqref{eq:CPconds}, as detailed in
Sec.~B
of~\cite{sm}, where we also present experimental results for this model.

In Fig.~\ref{fig:interf_scheme_sing_phot} we also show the optimal detection scheme that saturates the QFI, formally a projective measurement on the eigenstates of the SLD $L_{\theta_0}$:
$\frac{1}{\sqrt{2}} \ket{1} \pm \I \frac{e^{ -\I \theta_0}}{2} \left( \ket{2} + \ket{3} \right)$ and $\frac{1}{\sqrt{2}}\left( \ket{2} - \ket{3} \right)$.
Here $\theta_0$ is the working point around which we estimate small variations of the parameter.
Setting $\theta_0=0$ for concreteness, the probabilities of registering the photon at one of the three detectors read:
\begin{equation}
\label{eq:probAver1}
\begin{aligned}
p_1 &= \frac{1-p_3}{2} + f_\theta, &f_{\theta} &= v \eta \sqrt{q(1-q)} \sin\theta , \\
p_2 &= \frac{1-p_3}{2}-f_\theta &p_3 &= \frac{1}{2} (1-q) \left(1- v \kappa \right), 
\end{aligned}
\end{equation}
where we have also introduced an additional visibility parameter $0 \leq v \leq 1$ that takes into account imperfect interference---see Sec.~C of~\cite{sm} for a complete derivation.

If $v=1$ and $\kappa=1$, detector `3' never clicks ($p_3=0$).
This corresponds to perfectly correlated phases fluctuations, $c=1$, and results in an effective standard qubit dephasing model (modes $2$ and $3$ can be regarded as a single mode). When $\kappa$ is decreased, which corresponds to moving from the perfectly correlated noise ($c=1$) via  uncorrelated ($c=0$) to the perfectly anti-correalted one ($c=-1$), detector `3' filters-out the `anti-correlated' part of the phase noise.  Doing so, it effectively increases the  visibility of the interference pattern observed in detectors `1' and `2'
(note that the $\theta$ dependent `fringe-terms' $f_\theta$ in $p_1$ and $p_2$ do not depend on $\kappa$, see Eq.~\eqref{eq:probAver1}):
\begin{equation}
\mathcal{V} = \frac{p\left(1|\frac{\pi}{2} \right)-p\left(1|\frac{-\pi}{2} \right)}{p\left(1| \frac{\pi}{2} \right)+p\left(1|\frac{-\pi}{2} \right)} = \frac{4 v \eta \sqrt{q(1-q)} }{1+q + v \kappa (1-q)},
\label{eq:visibility}
\end{equation}
 and leads to a better phase sensitivity.
 The possibility of filtering out the anti-correlated part of the noise is the essence of the counter-intuitve behaviour of the QFI.

\emph{Experimental results.}
We have realized the proposed model in a photonic experiment that is schematically depicted in Fig.~\ref{fig:interf_scheme_sing_phot}.
The two reference arms of the three-arm interferometer are piezo-actuated which allow us to control both the phase to be measured $\theta$ and the zero-mean fluctuations $\phi_1$ and $\phi_2$, (see 
Sec.~D
of~\cite{sm} for a detailed description of the physical implementation).
The interferometer outputs are coupled to single mode fibers connected to superconducting nanowire single photon detectors (SNSPD) with coincidence counters.
We feed the interferometer with heralded single-photons at 795 nm generated in a continuously driven non-degenerate (795 nm and 805 nm) SPDC source.
We run the experiment in 80-second-long intervals and accumulate coincidences for various parameter settings $\{\theta, \phi_0, c\}\in[0,2\pi]\times[0,1.16]\times[-1,1]$. To emulate the discrete $\pm\phi_0$ phase kicks we measure each $\{\phi_1,\phi_2\}$ sub-setting separately and combine the results according to the given $c$.
The compact form of the interferometer provides sufficient 15-minutes-long phase stability.
To maintain long-time stability, each measurement interval is preceded by calibration step consisting of scanning the actuators and observing intensity fringes (without heralding) at the interferometer outputs. The calibration step not only allows us to stabilize $\theta$ value but also provides information about the intrinsic visibility $v\approx0.97$ and relative efficiencies $\chi$ in the three measurement ports $\chi(p_2/p_1)\approx0.95$, $\chi(p_3/p_1)\approx1.61$.
\begin{figure}[t]
\includegraphics[width=1\columnwidth]{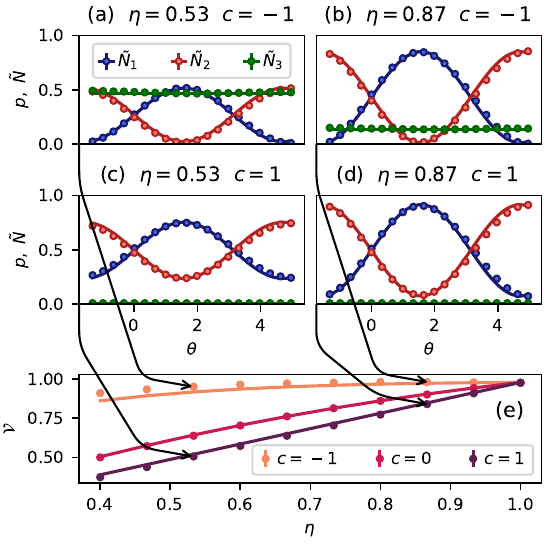}\caption{(a-d) Exemplary probability distributions for two different $\eta$ settings in perfectly correlated and anti-correlated case. The points correspond to measured and renormalized coincidences ($\tilde{N}$). (e) Measured (points) and expected (solid lines) visibility parameter for chosen $\eta$ and $c$ parameters.
\label{fig:p_vis}}
\end{figure}
We first compare the measured coincidences distributions with the theoretical predictions given by Eq.~\eqref{eq:probAver1}.
To recover the probabilities from measured counts we normalize the number of detected photons accounting for different net efficiencies at the interferometer output ports.
In Fig.~\ref{fig:p_vis}(a-d) we show examples of normalized counts $\tilde{N}$ measured for range of $\theta$ phases along with the expected curves corresponding to the same $\eta$ and $c$ parameters.
From those 
we recover the visibilities that are shown on a common plot in Fig.~\ref{fig:p_vis}(e).
The solid lines represent the expected behavior as described by Eq.~\eqref{eq:visibility}. 

\begin{figure}[t]
\includegraphics[width=1\columnwidth]{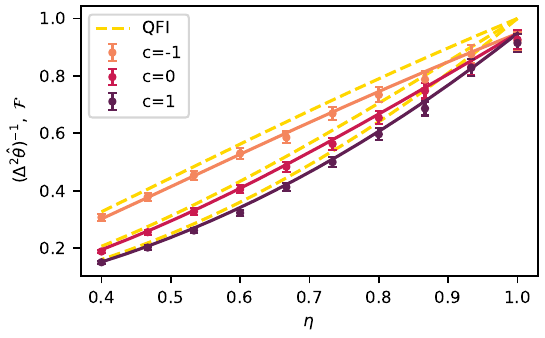}\caption{Precision of $\theta$ estimation (points) for set of $\eta$ and $c$ parameters compared with CRB and QCRB given by FI (solid lines) and QFI respectively.\label{fig:est}}
\end{figure}

In order to validate the improvement of the phase estimation precision 
in presence of increasingly anti-correlated noise, we have 
implemented estimation procedure based on the efficient 
locally unbiased estimator (at the working point $\theta_0=0$): 
\begin{equation}
    \hat{\theta}=\frac{N_{1}-N_{2}}{2v\eta\sqrt{(1-q)q}(N_{1}+N_{2}+N_{3})}
\end{equation} 
where $N_{i}$ denotes the detected photon counts in the given output port. 
We evaluate the estimator on $10^4$ randomly prepared sets of measurement outcomes each containing $10^5$ samples in total. Each set is prepared by randomly sampling (with repetitions) from the $10^5$ measured single-photon counts at the $\theta_0=0$ point.
The resulting single-photon estimation precision, defined as the inverse of the estimator variance divided by the total number of detected photons, is presented in Fig.~\ref{fig:est}. The errorbars correspond to chi-squared 99\% confidence interval. The rescaling to single-photon value allows us to compare the achieved precision with the classical FI (solid lines), calculated for the model probabilities (Eq.~\eqref{eq:probAver1})
that take into account the finite interference visibility $v$. The dashed lines represent the ultimate bound given by the QFI.
In Fig.~\ref{fig:est} we clearly see that the estimation precision improves for the anti-correlated noise as predicted by the FI and QFI. 

\emph{Entanglement.}
A similar effect can also be observed for entanglement. Let us apply the channel $\mathcal{E}$ to one part of the entangled state $\ket{\psi}^{AB} = \sqrt{q} \ket{11} + \sqrt{\frac{1-q}{2}}(\ket{22} + \ket{33})$. The resulting state takes the form of a maximally correlated state~\cite{RainsPhysRevA.60.179}:
\begin{align}
    \rho^{AB} & =\openone\otimes\mathcal{E}(\ket{\psi}\!\bra{\psi}^{AB}) = \sum_{i,j} \alpha_{ij}\ket{ii}\!\bra{jj} \label{eq:MC}
\end{align}
with parameters $\alpha_{11}=q$, $\alpha_{22}=\alpha_{33} = (1-q)/2$, $\alpha_{12}=\alpha_{21}=\alpha_{13}=\alpha_{31}=\eta \sqrt{q(1-q)/2}$, and $\alpha_{23}=\alpha_{32}=\kappa (1-q)/2$.
In fact, these states can be regarded as the maximally correlated states associated to the qutrit states in Eq.~\eqref{eq:quditDeph12}.
It is known that coherence properties of a quantum state are closely related to entanglement properties of the corresponding maximally correlated state~\cite{Streltsov2016a,WinterPhysRevLett.116.120404}.
Following this analogy, it is reasonable to expect that the results presented above will also extend to entanglement of the states in Eq.~\eqref{eq:MC}.
For this, we investigate distillable entanglement of $\rho^{AB}$, characterizing how many singlets can be extracted from the state $\rho^{AB}$ via local operations and classical communication in the asymptotic limit~\cite{BennettPhysRevLett.76.722,PlenioQuant-ph/0504163}.
While the distillable entanglement is hard to evaluate in general, a closed expression exists for states of the form~\eqref{eq:MC}: $E_\mathrm{d}(\rho^{AB}) = S(\rho^B) - S(\rho^{AB})$ with the von Neumann entropy $S(\rho) = -\Tr (\rho \log_2 \rho )$.
For this class of states only the the global entropy $S(\rho)$ depends on $\kappa$ and the distillable entanglement $E_\mathrm{d}(\rho^{AB})$ is equal to the relative entropy of coherence of the state in Eq.~\eqref{eq:quditDeph12} with respect to the orthonormal basis $\{\ket{1},\ket{2},\ket{3} \}$~\cite{Streltsov2016a,WinterPhysRevLett.116.120404}.
We can intuitively think that there are two competing effects at play: reducing $\kappa$ decreases the coherence in the subspace spanned by $\ket{2}$ and $\ket{3}$, but at the same time increases coherence between this subspace and $\ket{1}$.
For this reason, the effect is far less pronounced than with the QFI, but there is still a region of parameters where the entanglement increases when moving from a positive value of $\kappa$ to a smaller (yet positive) value.
More details on the entanglement analysis are presented in
Sec.~F
of~\cite{sm}.

\emph{Classical light.}
It is natural to wonder what would happen if instead of a single photon we used classical light, i.e. a coherent input state $\ket{\alpha}$.
In this case we use a Gaussian phase fluctuation model $P(\phi_1,\phi_2)$, 
see details in 
Sec.~B
of~\cite{sm}. 
We consider an estimation strategy, analogous to the standard MZ interferometer, based on the photon number difference $I^{-} = n_1 - n_2$, optimal in the single-photon regime, and we evaluate the variance using error propagation: $\Delta^2 \theta  = \Delta^2 I^{-} / \left| \frac{d }{ d \theta} \mathbb{E}\left[\langle I^{-} \rangle\right] \right|^2$, where the $\Delta^2 I^{-}$ includes also the contribution of the distribution $P(\phi_1,\phi_2)$ of the fluctuating reference phases.
In 
Sec.~D
of~\cite{sm} we show that
\begin{equation}
\label{eq:cohVarTheta}
  \Delta^2 \theta   = \frac{1}{|\alpha|^2 \mathcal{F}_{\theta} } + 
   \frac{1}{2} \left( \eta^{-2} - \eta^2 + \eta^{-2 c} - \eta^{2 c}  \right),
\end{equation}
where the first terms contains the single-photon QFI~\eqref{eq:qutritQFIetakappa} (with the $\eta$ and $\kappa$ now related to the parameters $\sigma$ and $c$ of the Gaussian noise model, see
Eq.~(B2)
in~\cite{sm}).
This result implies that this strategy with classical light performs worse than repeating a single photon experiment many times, but for a weak coherent state we have the same sensitivity per average photon, i.e. $(|\alpha|^2 \Delta^2 \theta)^
{-1} \stackrel{\alpha \to 0}{\longrightarrow} \mathcal{F}_\theta$.
For an intense beam the variance saturates to a constant value, as in standard phase diffusion~\cite{Escher2012,Macieszczak2014}, and only vanishes for $c=-1$.

Finally, we can also consider a `naive classical' strategy, without any interference between the two reference modes, which is equivalent to two separate MZ interferometers sensing the phase shifts $\theta-\phi_1$  and $\theta-\phi_2$.
In this case one finds the same structure as in $\eqref{eq:cohVarTheta}$, but the first term corresponds to the QFI evaluated at $c=1$, i.e. it does not improve by reducing the correlations.
Details are given in 
Sec.~G.2
of Ref.~\cite{sm}.

\emph{Discussion.}
Starting with an abstract model of the loss of coherence within a degenerate subspace of the generator of a phase shift, we have discussed and demonstrated the nonintuitive effect of an increased sensing precision.
By considering the optimal interferometric scheme we have related such an improvement to the possibility of filtering out the anti-correlated part of the dephasing noise.
This effect may be of practical importance in interferometric schemes where one is able to cause the phase fluctuations affecting multiple reference modes to become uncorrelated (by e.g. separating them spatially)  or even make them anticorrelated (if thermal fluctuations were responsible for dephasing, while the materials had opposite thermal and the temperature coefficient of the refractive index).
Interestingly, the most significant gains from the effect can be obtained at the single photon level, and when the light from both the reference arms is interfered with each other before eventually interfering with the light from the signal arm.
The results can be generalized to obtain qualitatively similar effects in case of $d$ reference modes instead of just $2$.

\emph{Acknowledgments.}
We thank Konrad Banaszek, Wojtek Górecki and Wojciech Wasilewski for fruitful discussions.
F.~A. and R.~D.~D. were supported by National Science Center (Poland) Grant No. 2016/22/E/ST2/00559.
R.~D.~D. was additionally supported by the National Science Center (Poland) grant No.\ 2020/37/B/ST2/02134.
M.~M., M.~L., A.~S. and M.~P. are supported by the Foundation for Polish Science “Quantum Optical Technologies” project (MAB/2018/4), carried out within the International Research Agendas programme of the Foundation for Polish Science co-financed by the European Union under the European Regional Development Fund.

\nocite{VedralPhysRevLett.78.2275,Rains959270,Zinchenko2010,Girard2015a}

\bibliography{bibliographyPRL}

\clearpage
\appendix
\setcounter{figure}{0}
\setcounter{page}{1}
\setcounter{equation}{0}
\renewcommand{\thefigure}{S\arabic{figure}}
\renewcommand{\theequation}{S\arabic{equation}}
\renewcommand{\appendixname}{}

\begin{center}
	\textbf{\large Supplemental Material}
\end{center}

The Supplemental Material is organized as follows.
In Sec.~\ref{app:details} we give a few more details on the calculations performed to obtain the results in the main text.
In Sec.~\ref{app:Gaussian} we study the case of Gaussian phase fluctuations, mentioned in the main text, for which we have also performed an experiment.
In Sec.~\ref{app:visibility} we derive detector clicks probabilities in the three-arm interferometer taking into account imperfect visibility.
In Sec.~\ref{app:implementation} we give more details on the experimental implementation of the interferometer.
In Sec.\ref{app:variance} we provide full statistics of obtained estimated values of the phase in case of both discrete and Gaussian noise models.
In Sec.~\ref{app:entanglement} we present the results mentioned in the main text about the behaviour of entanglement as a function of $\kappa$.
In Sec.~\ref{app:classical} we give more details on the calculations for classical light.

\section{Details on the calculations}
\label{app:details}

The state after the parameter is encoded is
\begin{equation}
  \begin{split}
 & \rho_{\theta} = \mathcal{E}_{\theta}(\rho) = e^{-\I \theta G} \mathcal{E}(\rho) e^{\I \theta G} = \\
  &  \begin{bmatrix}
     q & \frac{\eta  \sqrt{q(1-q)} e^{-\I \theta  \Delta}}{\sqrt{2}} & \frac{\eta  \sqrt{q(1-q)} e^{ -\I \theta  \Delta }}{\sqrt{2}} \\
     \frac{\eta  \sqrt{q(1-q)} e^{ \I \theta \Delta}}{\sqrt{2}} & \frac{1-q}{2} & \frac{1}{2} \kappa  (1-q) \\
     \frac{\eta  \sqrt{q(1-q)} e^{ \I \theta \Delta}}{\sqrt{2}} & \frac{1}{2} \kappa  (1-q) & \frac{1-q}{2} \\
    \end{bmatrix},
  \end{split}
\end{equation}
where $\Delta = E_1-E_2$.
The purity of this state is 
\begin{equation}
  \Tr [ \rho_{\theta}^2 ]  = \frac{1}{2} \left(\kappa ^2+q \left(4 \eta ^2-4 \eta ^2 q+\kappa ^2 (q-2)+3 q-2\right)+1\right),
\end{equation}
which is an increasing function of $\kappa$ when $\kappa > 0$.

The QFI $\mathcal{F}_{\theta}(\rho)=\Tr [ \rho_\theta L_{\theta}^2 ]$ is defined in terms of the Hermitian symmetric logarithmic derivative (SLD) operator $L_{\theta}$ that satisfies the equation $2 \frac{d \rho_\theta}{d \theta} = L_{\theta} \rho_{\theta} + \rho_{\theta} L_{\theta}$.
For this problem the SLD is 
\begin{align}
  L_\theta &= \frac{2}{1 + q (1-\kappa) + \kappa} \frac{d \rho_{\theta} }{d\theta} \\
  &= 
  \frac{ \eta  \left( 2 \Delta \sqrt{2 (1-q) q } \right) }{q(1-\kappa) + 1 + \kappa } 
  \begin{bmatrix}
     0 & -\I e^{- \I \Delta  \theta } & -\I e^{-\I \Delta  \theta }  \\
     \I e^{ \I \Delta  \theta }& 0 & 0 \\
     \I e^{ \I \Delta  \theta }& 0 & 0 \\
    \end{bmatrix},
\end{align}
which is simply a rescaling of a factor $\eta  /\left[q(1-\kappa) + 1 + \kappa \right]$ of the pure-state SLD $ 2 \left(\ket{d \psi_\theta / d\theta }\bra{ \psi_\theta } + \ket{\psi_\theta }\bra{ d  \psi_\theta / d\theta} \right)$ which would be obtained for $\eta = \kappa = 1$.

\section{Gaussian fluctuations}
\label{app:Gaussian}
Here we treat the phases $\phi_1$ and $\phi_2$ as zero-mean correlated Gaussian random variables with the following bivariate normal density
\begin{equation}\label{eq:bivarGauss}
p( \phi_1 , \phi_2 ) = \frac{1}{2 \pi \sigma^2 \sqrt{1-c^2}} \exp \left[ -\frac{1}{2} \boldsymbol{\phi}^T
\begin{pmatrix} \sigma^2 && c \sigma^2 \\ c \sigma^2 && \sigma^2  \end{pmatrix}^{-1} \boldsymbol{\phi}  \right],
\end{equation}
where $\boldsymbol{\phi}^T=[\phi_1,\phi_2]$, the standard deviation $\sigma > 0$ is equal for both variables and where $c=\frac{\mathbb{E}[\phi_1 \phi_2]}{\sqrt{\mathbb{E}[\phi_1^2]\mathbb{E}[\phi_2^2]}}$ is the standard correlation coefficient, satisfying $-1 \leq c \leq 1$.
The two limiting values $c=\pm 1$ correspond to perfect correlations, thus the covariance matrix is singular, however it is safe to take the limit $c \to \pm 1$ at the end of the calculations of the quantities of interest.
The parameters of the qutrit channel are related to the parameters of this distributions as follows:
\begin{equation}
\label{eq:etakappaGauss}
\eta = e^{-\frac{\sigma ^2}{2}} , \, \kappa = e^{-\sigma^2 (1-c)},
\end{equation}
inverted to obtain
\begin{equation}
\sigma^2 = - 2 \log \left(\eta\right)
\qquad
c = 1-\frac{\log \kappa}{2 \log \eta}.
\end{equation}
from which we see that the dephasing parameter $\kappa$ cannot span the whole physicality region (Eq.~(2) in the main text).
As we go from $c=-1$ (perfectly anti-correlated fluctuations) to $c=1$ (perfectly correlated) the parameter $\kappa$ spans the range $\eta^4\leq \kappa \leq 1$.
In other words, going from anti-correlated to correlated fluctuations we go from strong to no dephasing in the subspace $\mathrm{span}(\ket{2},\ket{3})$ and in particular we can never achieve a negative $\kappa$.

\begin{figure}[H]
\includegraphics[width=1\columnwidth]{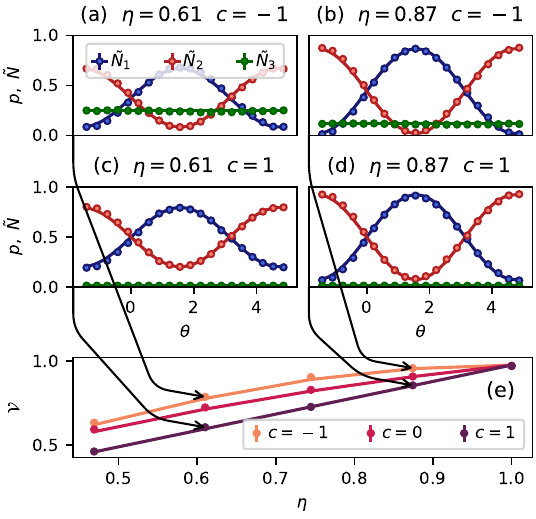}
\caption{(a-d) Exemplary probability distributions for two different $\eta$ settings for perfectly correlated and anti-correlated Gaussian noise. The points correspond to measured and renormalized coincidences ($\tilde{N}$). (e) Measured (points) and expected (solid lines) visibility parameter for chosen $\eta$ and $c$ parameters.   \label{fig:p_vis_gaussian}}
\end{figure}

\begin{figure}[H]
\includegraphics[width=1\columnwidth]{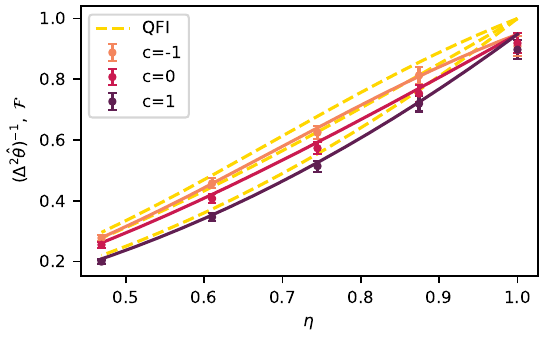}\caption{Precision of $\theta$ estimation (points) for set of $\eta$ and $c$ parameters compared with CRB and QCRB given by FI (solid lines) and QFI respectively in case of Gaussian fluctuations.\label{fig:gaussian_est}}
\end{figure}

The Gaussian model has been also realized experimentally. For this the piezo-actuated arms were directly programmed to move according to numerically generated Gaussian noise with Gaussian power spectrum of width smaller than the piezo-actuator bandwidth. The results analyzed in the same manner as in the case of discrete noise are presented in Fig. \ref{fig:p_vis_gaussian}. The experimental data are in good agreement with the theoretical predictions and the visibility improvement for anti-correlated noise is clearly visible. Finally, in Fig. \ref{fig:gaussian_est} we show $\theta$ estimation results compared with the FI and QFI.

\section{Detection probabilities in the three-arm interferometer with imperfect visibility}
\label{app:visibility}
Let us first derive probabilities of clicks of the three detectors clicks in Fig.~1 in the main text 
in an idealized scenario of perfect modes matching an hence perfect interference. 
We introduce a notation, where a beam-splitter with power transmission $q$ transforms input state amplitudes according to the following matrix: 
\begin{equation}
\textrm{BS}_q =  \begin{bmatrix}
  \sqrt{1-q} & - \sqrt{q}\\
  \sqrt{q} & \sqrt{1-q}
  \end{bmatrix}.
\end{equation}
Let us denote the the state of a single photon just before the detection  $[\psi_1,\psi_2,\psi_3]^T$, where $\psi_i$ is the amplitude of the photon being in the $i$-th arm of the interferometer (arms numbered in agreement with the labels of the respective detectors)
\begin{multline}
\label{eq:ampli}
\begin{bmatrix}
\psi_1\\ \psi_2\\ \psi_3
\end{bmatrix} = 
 \textrm{BS}^{(12)}_{50\%} \cdot  \textrm{BS}^{(23)}_{50\%} \cdot \begin{bmatrix}
 e^{- i \theta} & 0  & \\
 0 &  e^{-i \phi_1 - i \frac{\pi}{2}} & \\
 0 & 0 &  e^{-i \phi_2 + i \frac{\pi}{2}}
 \end{bmatrix} 
 \cdot \\
 \cdot \textrm{BS}^{(23)}_{50\%}  \cdot \textrm{BS}^{(12)}_{q}  \cdot 
 \begin{bmatrix} 0 \\ 1 \\0  \end{bmatrix}  =  
 \begin{bmatrix} \sqrt{\frac{q}{2}}e^{-i \theta} - i \sqrt{\frac{1-q}{8}}\left(e^{- i \phi_1} + e^{-i \phi_2} \right) \\ 
 -\sqrt{\frac{q}{2}}e^{-i \theta} - i \sqrt{\frac{1-q}{8}}\left(e^{- i \phi_1} + e^{-i \phi_2} \right)
 \\ \sqrt{\frac{1-q}{4}} i\left( e^{-i \phi_1} - e^{- i \phi_2}  \right)  \end{bmatrix}
\end{multline}
where the superscripts in $\textrm{BS}_q^{(ij)}$ indicate the modes on which a given beam-splitter acts. Probabilities of detector clicks are then computed as
\begin{equation}
\label{eq:probideal}
p^{\textrm{ideal}}_i = \mathbb{E}\left[|\psi_i|^2\right], 
\end{equation}
where $\mathbb{E}\left[ \circ \right]$
represents averaging with respect to fluctuating phases, while superscript `ideal' indicates that the probability corresponds to the ideal interferometer with perfect visibility. Recalling that 
$\eta = \mathbb{E}[e^{i \phi_i} ]$, 
$\kappa = \mathbb{E}[e^{i (\phi_1-\phi_2)}]$, and plugging Eqs.~\eqref{eq:ampli} into Eq.~\eqref{eq:probideal} we arrive at:
\begin{align}
p_1^{\textrm{ideal}} & =
 \frac{1}{2} \left[  1 -p_3^{\textrm{ideal}}\right] + \eta \sin \theta \sqrt{q(1-q)},
\\
p_2^{\textrm{ideal}} & = \frac{1}{2} \left[  1 - p_3^{\textrm{ideal}}\right] - \eta \sin \theta \sqrt{q(1-q)},\\
p_3^{\textrm{ideal}} & = \frac{(1-q)(1-\kappa)}{2}. 
\end{align}

There are a number of physical factors that prevent light beams from interfering perfectly:  mode-mismatch, unequal losses in the interferometer's arms and imbalances caused by the non-ideal beam-splitter implementation.  We will model this situation in the simplest possible way using a single visibility parameter $0 \leq v \leq 1$.
The final probabilities will be computed according to the following rule:
\begin{equation}
    p_i = v p_i^{\textrm{ideal}} + (1-v) p_i^{\textrm{incoh}} 
\end{equation}
where $p_i^{\textrm{incoh}}$ represent probabilities of detector clicks in case no interference phenomena were present and
the light impinging on the beam-splitters was completely incoherent:
\begin{equation}
p_1^{\textrm{incoh}} = p_2^{\textrm{incoh}} = \frac{1+q}{4}, \ p_3^{\textrm{incoh}} = \frac{1-q}{2}.
\end{equation}
As a reuslt we obtain 
\begin{align}
p_1 &= \frac{1-p_3}{2} + f_\theta, &f_{\theta} &= v \eta \sqrt{q(1-q)} \sin\theta , \\
p_2 &= \frac{1-p_3}{2}-f_\theta &p_3 &= \frac{1}{2} (1-q) \left(1- v \kappa \right), 
\end{align}
as reported in the main text. This model proved sufficient to fit all the data obtained in the experiment.

\section{Three arm interferometer implementation}
\label{app:implementation}
\begin{figure}[t]
\includegraphics[width=1\columnwidth]{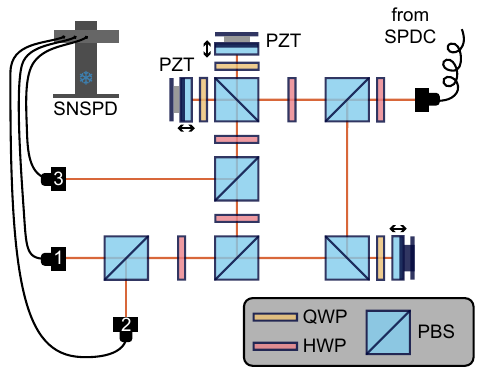}
\caption{Implementation of the three arm interferometer. PBS - polarization beam splitter, HWP - half waveplate, QWP - quarter waveplate, SNSPD - superconducting nanowire single photon detector (ID Quantique ID281).}
\label{fig:interf_true_scheme}
\end{figure}
The interferometer is built in a free-space MZ configuration with a Michelson-like polarization interferometer incorporated in one of the arms as depicted in Fig. \ref{fig:interf_true_scheme}. The two Michelson arms are piezo-actuated (PZT), which allow us to control both the phase to be measured $\theta$ and the zero-mean fluctuations $\phi_1$ and $\phi_2$. The optimal state preparation is achieved by rotating the half-waveplate (HWP) at the interferometer input that virtually changes the first (polarization) beam-splitter transmittance $q$.  As the SPDC photons have relatively short wavepackets (ca. 100 fs), we calibrate the delay of the heralding photon (at 805 nm) and we set all three arms to be equally long to guarantee good interference visibility. The quality of single photons used in the experiment was validated by heralded second-order auto-correlation measurement $g^{(2)}=\langle{n(n-1)}\rangle/\langle n\rangle^2=0.00064\pm0.00002$.

\section{Variance estimation}
\label{app:variance}
\begin{figure*}[t]
\includegraphics[width=1\textwidth]{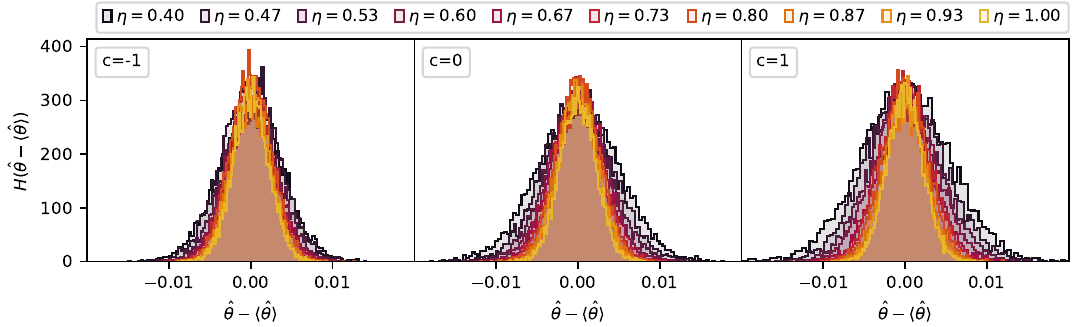}
\caption{Centered histograms of $\theta$ estimation results for the case of discrete phase kicks. Binnings of the histograms are chosen to yield 100 bins in each case.}
\label{fig:dstat}
\end{figure*}

\begin{figure*}[t]
\includegraphics[width=1\textwidth]{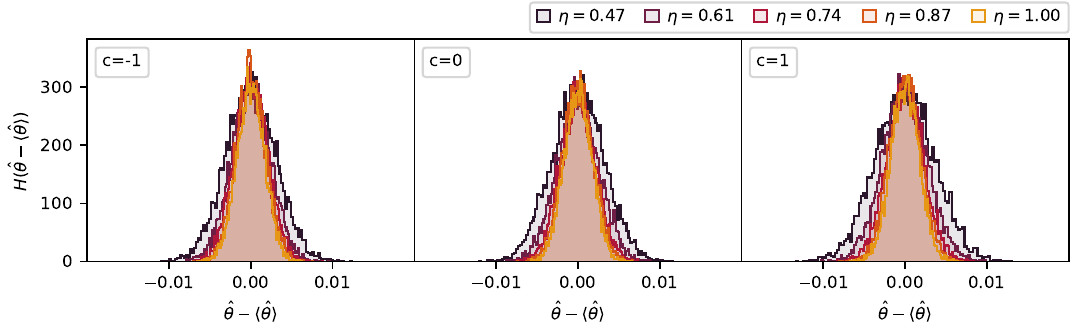}
\caption{Centered histograms of $\theta$ estimation results for the case of Gaussian phase fluctuations. Binnings of the histograms are chosen to yield 100 bins in each case.}
\label{fig:gstat}
\end{figure*}

Here we provide statistical data obtained from $10^4$ random experiments to estimate the variance of the $\hat{\theta}$ estimator in both discrete Fig.~(\ref{fig:dstat}) and Gaussian noise Fig.~(\ref{fig:gstat}) cases. Each histogram contains $10^4$ estimates of $\hat{\theta}$ binned into 100 equal-width intervals. The variance of the estimator is simply inferred from the distribution. Assuming that the $n$ samples come from normal distribution with variance $\sigma_{\hat{\theta}}^2$, the uncertainty of the variance estimate can be derived from chi-squared test as $\frac{(n-1)\Delta^2\hat{\theta}}{\sigma_{\hat{\theta}}^2}$ has a chi-squared distribution. For $n=10^4$ samples the 99\% confidence interval for $\Delta^2\hat{\theta}$ is between $0.965$ and $1.037$ of nominal value.


\section{Entanglement}
\label{app:entanglement}

For a bipartite state $\rho^{AB}$ the relative entropy of entanglement is defined as~\cite{VedralPhysRevLett.78.2275}
\begin{equation}
E_\mathrm{r}(\rho^{AB})=\min_{\sigma\in\mathcal{S}}S(\rho||\sigma)
\end{equation}
with the quantum relative entropy $S(\rho||\sigma) = \Tr[\rho \log_2 \rho] - \Tr[\rho \log_2 \sigma]$, and $\mathcal{S}$ is the set of separable states. For maximally correlated states the relative entropy of entanglement coincides with the distillable entanglement, and both are given as~\cite{Rains959270}
\begin{equation}
E_\mathrm{d}(\rho^{AB})=E_\mathrm{r}(\rho^{AB})=S(\rho^{B})-S(\rho^{AB}). \label{eq:EdMC}
\end{equation}

\begin{figure*}[th]
  \centering
  \begin{tikzpicture}
    \node (baz) at (0pt,0pt) {\includegraphics[width=0.33\textwidth]{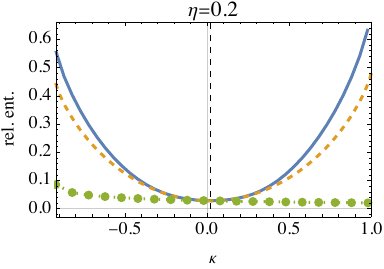}};
    \node [anchor=north west] at (baz.north west) {\textbf{(a)}};
    \node (bad) at (170pt,0) {\includegraphics[width=0.33\textwidth]{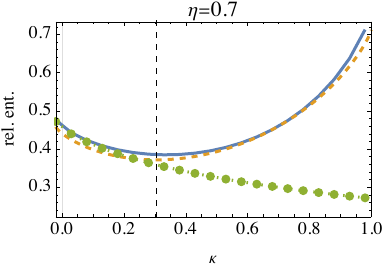}};
    \node [anchor=north west] at (bad.north west) {\textbf{(b)}};
    \node (bag) at (340pt,0) {\includegraphics[width=0.33\textwidth]{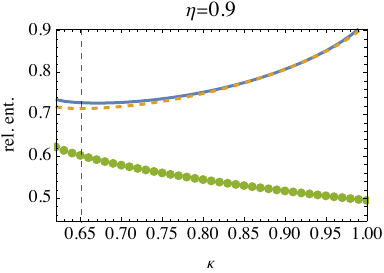}};
    \node [anchor=north west] at (bag.north west) {\textbf{(c)}};
  \end{tikzpicture}
  \caption{Relative entropy of entanglement.
  Solid blue: optimal state (optimized over $q$), dashed orange: maximally entangled states ($q=1/3$), green markers: $\frac{1}{\sqrt{2}}\left[ \ket{11} + \frac{1}{2} (\ket{2}+\ket{3})\otimes(\ket{2}+\ket{3})  \right] $.
  The vertical line is drawn at the minimum of the curve for maximally entangled states, showing that in the region at its left entanglement increases by having a more dephased state.} 
  \label{fig:entanglement}
\end{figure*}

In Fig.~\ref{fig:entanglement} we plot the relative entropy of entanglement for states of the form 
\begin{align}
    \rho^{AB} & =\openone\otimes\mathcal{E}(\ket{\psi}\!\bra{\psi}^{AB})
\end{align}
with $\ket{\psi}^{AB} = \sqrt{q} \ket{11} + \sqrt{\frac{1-q}{2}} \left( \ket{22} + \ket{33} \right)$, which we conjecture to be optimal (optimizing $q$) and for $q=1/3$ are maximally entangled. In this case $\rho^{AB}$ is maximally correlated, which means that we can evaluate the relative entropy of entanglement and distillable entanglement analytically via Eq.~(\ref{eq:EdMC}).

In Fig.~\ref{fig:entanglement} we also consider states of the form \begin{equation}
    \ket{\psi}^{AB} = \frac{1}{\sqrt{2}}\left[ \ket{11} + \frac{1}{2} (\ket{2}+\ket{3})\otimes(\ket{2}+\ket{3})  \right].
\end{equation}
In this case  the relative entropy of entanglement~\cite{VedralPhysRevLett.78.2275},
 evaluated numerically using the algorithm of~\cite{Zinchenko2010,Girard2015a}, always increases as $\kappa$ decreases from one to zero.

Additionally, results corresponding to maximally entangled states are included in the plot, $q=1/3$, and appear to be optimal for a uniformly dephasing channel, i.e., $\kappa=\eta$.

\section{Classical light with Gaussian fluctuations}
\label{app:classical}

The theoretical and experimental results exposed in the main text focus on a three-arm interferometer operated with single-photon input states, abstractly described as a qutrit.
A simple and natural generalization of this scheme is to send classical light, i.e. a coherent state, instead of a single photon state.
In this scenario noise models with a finite number of discrete phase kicks are not well suited and it more realistic to assume the Gaussian noise introduced in Sec.~\ref{app:Gaussian}.
In this Section we present a cursory analysis of the metrological performances of classical light and Gaussian fluctuations.

The analytical evaluation of the QFI of the quantum state at the output or the FI of the particular measurement scheme is much harder in this case.
For this reason we fix a particular intuitive estimation strategy and we resort to a simpler error propagation analysis.
Overall we show that operating the interferometric setup considered in the main text with single photons is a better way to take advantage of partially anti-correlated phase fluctuations than a ``semiclassical'' approach based on photon number differences.

We have also performed a preliminary numerical analysis that shows a different overall picture when considering the QFI as figure of merit.
However, we have numerical evidence that the estimation precision identified by the QFI is not attained by any linear interferometer with photon counting and would thus require nonlinear optical elements, making it far less amenable to practical implementations.

\subsection{Standard three arm interferometer}

We consider the three arm interferometer with random reference phases $\phi_1$ and $\phi_2$ and signal phase $\theta$.
Using classical light, the photon counts at the three outputs are independent and described by three Poissonian random variables with mean
\begin{align}
\langle N_1 \rangle =& \frac{\bar{n}_0}{4}  \biggl\{ 1 + q + (1-q) \cos ( \phi_1 - \phi_2 ) \\ 
&+ 2 \sqrt{ q ( 1- q)} \left[ \sin(\theta-\phi_1) + \sin(\theta -\phi_2) \right] \biggr\} \nonumber \\  
\langle N_2 \rangle =&  \frac{\bar{n}_0}{4}  \biggl\{ 1 + q + (1-q) \cos ( \phi_1 - \phi_2 ) \\ 
& - 2 \sqrt{ q ( 1- q)} \left[ \sin(\theta-\phi_1) + \sin(\theta-\phi_2) \right] \biggr\} \nonumber \\
\langle N_3 \rangle =& \frac{\bar{n}_0}{2} (1-q) \left[ 1 - \cos ( \phi_1 - \phi_2) \right],
\end{align}
where the expectation value in angular brackets is taken with respect to the initial coherent state of energy $|\alpha|^2 = \bar{n}_0$

To estimate the phase we consider the observable $I^{-} = N_1 - N_2$, corresponding essentially to a standard interferometric scheme.
For fixed values of the phases $\phi_1$ and $\phi_2$, this observable has the following mean and variance
\begin{align}
\bar{I}^{-}(\phi_1,\phi_2) &= \langle I^{-} \rangle = \langle N_1 \rangle - \langle N_2 \rangle \label{eq:Idifphi12} \\
&= \bar{n}_0 \sqrt{ q ( 1- q)} \left[ \sin(\theta-\phi_1) + \sin(\theta-\phi_2) \right] \nonumber \\
\Delta^2 I^{-}(\phi_1,\phi_2) & = \langle I^{-2} \rangle - \langle I^{-} \rangle^2 = \langle N_1 \rangle + \langle N_2 \rangle \label{eq:DeltaIdifphi12}  \\
&= \frac{\bar{n}_0}{2}  \left[ 1 + q + (1-q) \cos ( \phi_1 - \phi_2 ) \right].  \nonumber
\end{align}
However, we assume that the values of $\phi_1$ and $\phi_2$ have random fluctuations drawn from the bivariate normal distribution~\eqref{eq:bivarGauss}; as we did previously, we denote the expectation value with respect to this classical distribution with the symbol $\mathbb{E}$.
The final average value for the chosen observable is therefore
\begin{equation}
\bar{I}^{-} = \mathbb{E} \left[ \bar{I}^{-}(\phi_1,\phi_2) \right]  = 
2 \bar{n}_0 \sqrt{q(1-q)} e^{-\frac{\sigma ^2}{2}} \sin \theta
, 
\end{equation}
while from the law of total variance we have
\begin{align}
\Delta^2 I^{-} 
 = \Delta_Q^2 I^{-} + \Delta_C^2 I^{-}, \nonumber
\end{align}
where we have a ``quantum'' part $\Delta_Q^2 I^{-} = \mathbb{E} \left[ \Delta^2 I^{-}(\phi_1,\phi_2) \right]$ and a ``classical'' part $\Delta_C^2 I^{-} = \mathbb{E} \left[ \bar{I}^{-}(\phi_1,\phi_2)^2 \right] - \mathbb{E}_{\phi} \left[ \bar{I}^{-}(\phi_1,\phi_2) \right]^2$.
More explicitly, these two terms have the following expressions
\begin{align}
\Delta_Q^2 I^{-} =& \frac{1}{2} \bar{n}_0 \left( (1-q) e^{(c-1) \sigma ^2} + q +1\right) \label{eq:QuantumErr}\\ 
\Delta_C^2 I^{-} =& \bar{n}_0^2 (1-q) q \biggl\{ 
1 + e^{(c -1) \sigma ^2} \\
& + e^{-2 \sigma ^2} \left[ \cos 2 \theta  \left(e^{(1-c) \sigma ^2}+1\right)+4 e^{\sigma ^2} \sin ^2 \theta \right] \biggr\}. \nonumber
\end{align}
We can notice that the classical term is quadratic in the input intensity $\bar n_0$ while the quantum is linear, therefore for very weak coherent light only the quantum contribution survives.
Moreover, around the optimal working point $\theta = 0$ the classical term vanishes $\Delta_C^2 I^{-} = 0$ for perfectly anticorrelated fluctuations ($c=-1$).

The quantity we want to compute is the variance  obtained through error propagation:
\begin{align}
\label{eq:VarThetaCoh}
& \Delta^2 \theta  = \frac{\Delta^2 I^{-}}{\left| \frac{d }{d \theta} \bar{I}^{-} \right|^2}
 = \frac{\Delta_Q^2 I^{-}}{\left|  \frac{d }{d \theta} \bar{  I}^{-} \right|^2} + \frac{\Delta_C^2 I^{-}}{\left|  \frac{d }{d \theta}\bar{I}^{-} \right|^2},
\end{align}
the quantum part is simply the single-photon QCRB rescaled by the average number of input photons:
\begin{equation}
\frac{\Delta_Q^2 I^{-}}{\left|  \frac{d }{d \theta}\bar{I}^{-} \right|^2} = \frac{1}{\bar{n}_0  \mathcal{F}_1 },
\end{equation}
where $\mathcal{F}_\theta$ is the single-photon QFI (Eq.~(3) in the main text) with $\eta$ and $\kappa$ obtained for Gaussian fluctuations~\eqref{eq:etakappaGauss}, meaning that for a very weak coherent state $\bar{n}_0 \ll 1$ the ``sensitivity per average photon'' $\left( \bar{n}_0 \Delta^2 \theta \right)^{-1} $ coincides with the single-photon QFI.
This also implies that
\begin{equation}
\Delta^2 \theta \geq \frac{1}{\bar{n}_0 \mathcal{F}_1},
\end{equation}
so we see that for this estimation strategy, repeating a single photon experiment is better than using the same average amount of photons in a classical beam.
 
The variance $\Delta^2 \theta$ reaches an asymptotic value for $\bar n_0 \to \infty$, analogously to phase diffusion in a two-arm interferometer~\cite{Escher2012,Demkowicz-Dobrzanski2015a}, since $\frac{\Delta_Q^2 I^{-}}{\left|  \frac{d }{d \theta} \bar{I}^{-} \right|^2}$ decreases as $1/{\bar n_0}$ but the second term is constant
\begin{equation}
\label{eq:asympErrorCoh}
\frac{\Delta_C^2 I^{-}}{\left| \frac{d }{d \theta} \bar{I}^{-} \right|^2} = \sinh\left[ (c+1) \frac{\sigma^2}{2} \right]\cosh\left[ (c-1) \frac{\sigma^2}{2} \right] ,
\end{equation}
and vanishes only for $c=-1$.
We also see that this term (and thus the asymptotic variance) does not depend on the transmissivity $q$ of the initial beam-splitter, but we need to assume $0<q<1$, otherwise if $q=0$ or $q=1$ the quantum term of the variance would diverge.

\subsection{Four-arm scheme without interference between reference modes}
\label{app:fourarm}
\begin{figure}[t]
\includegraphics[width=1\columnwidth]{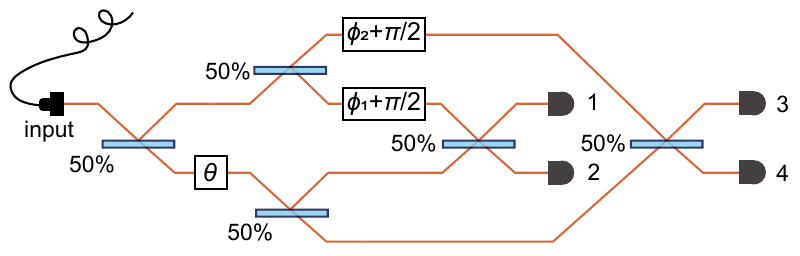}\caption{Alternative realization of the interferometer with separate, decoupled detection using both reference arms.\label{fig:3arm_intuitive}}
\end{figure}
It is quite natural to compare our three arm setup where we interfere coherently the two reference modes to a setup where each reference mode is interfered separately with the signal mode (see Fig. \ref{fig:3arm_intuitive}).
Practically, this means splitting the signal mode with a balance beam splitter after the phase is encoded.
However, this is essentially equivalent to using two separate MZ interferometers to sense two relative phase shifts $\theta - \phi_1$ and $\theta - \phi_2$, operating each of them with a coherent state with $\bar n_0 / 2$ average photons.
This approach yields the following average photon numbers at the output
\begin{align}
\langle N_1 \rangle  =& \frac{\bar{n}_0}{4} \left( 1  + 2\sqrt{q(1-q)}\sin(\theta-\phi_1)  \right) \\ 
\langle N_2 \rangle  =& \frac{\bar{n}_0}{4} \left( 1  - 2\sqrt{q(1-q)}\sin(\theta-\phi_1)  \right)\\
\langle N_3 \rangle  =& \frac{\bar{n}_0}{4} \left( 1  + 2\sqrt{q(1-q)}\sin(\theta-\phi_2)  \right) \\
\langle N_4 \rangle =& \frac{\bar{n}_0}{4} \left( 1  - 2\sqrt{q(1-q)}\sin(\theta-\phi_2)  \right).
\end{align}
To estimate the phase $\theta$ we consider the observable $I_{4}^{-} =  N_1 + N_3 - ( N_2 + N_4)$, corresponding to the sum of the signals of the two MZ interferometers, which recovers the same average value~\eqref{eq:Idifphi12} as in the three mode setup:
\begin{align}
\langle I_4^{-} \rangle &= \bar{n}_0 \sqrt{q(1-q)} \left[ \sin(\theta-\phi_1) +  \sin(\theta-\phi_2) \right] \label{eq:signal4modeCoh} \\
\Delta^2 I_4^{-} &=  \bar{n}_0,
\end{align}
but in this configuration we have a larger variance, which does not depend on the nature of the fluctuations.
The variance of the estimator constructed from this average corresponds again to~\eqref{eq:VarThetaCoh}, but now the quantum part is $\mathbb{E}\left[ \Delta^2 I_4^{-} \right] /\left|  \frac{d }{d \theta} \mathbb{E}\left[ \langle I_4^{-} \rangle \right] \right|^2 = \frac{1}{\bar{n}_0 \mathcal{F}_1}  \vert_{c=1}$, where $\mathcal{F}_1\vert_{c=1}$ is the single-photon QFI for perfectly correlated fluctuation.
Therefore without interfering the reference modes the scheme is unable to reap the benefits of anticorrelations.

Also in this scenario the conclusions would be different if one considered more complicated detection strategies and evaluated the QFI instead of the variance of a particular estimator with error propagation.
We leave a detailed study and comparison of various schemes based on strong classical light for the future works.


\end{document}